\documentclass[prd,twocolumn,floatfix,preprintnumbers,letterpaper]{revtex4}

\usepackage{epsfig}
\usepackage{latexsym,float}
\usepackage{graphicx,amssymb,amsbsy,bm,amsmath,psfrag,dsfont}
\usepackage{hyperref}
\usepackage[dvips]{color}

\newcommand{\be}{\begin{equation}}
\newcommand{\ee}{\end{equation}}
\newcommand{\ba}{\begin{array}}
\newcommand{\ea}{\end{array}}
\newcommand{\beq}[1]{\begin{equation}\label{#1}}
\newcommand{\eeq}{\end{equation}}
\newcommand{\bea}{\begin{eqnarray}}
\newcommand{\eea}{\end{eqnarray}}

\newcommand{\rf}[1]{(\ref{#1})}

\newcommand{\Det}{\rm{Det}}
\newcommand{\ustar}{u_\star}

\begin{document}

\title{Physical and Stable Closed Timelike Curves}
\author{Chiu Man Ho}
\email{chiuman.ho@vanderbilt.edu}
\affiliation{Department of
  Physics and Astronomy, Vanderbilt University, Nashville, TN 37235, USA}
\author{Thomas J. Weiler}
\email{tom.weiler@vanderbilt.edu}
\affiliation{Department of Physics and Astronomy, Vanderbilt University, Nashville, TN 37235, USA}
\date{\today}

\begin{abstract}
We construct a class of closed timelike curves (CTCs) using a compactified
extra dimension $u$. A nonzero metric element $g_{tu}(u)$ enables particles to travel backwards
in global time $t$. The compactified dimension guarantees that the geodesic curve closes in $u$.
The effective 2D ($t$ and $u$) nature of the metric ensures that spacetime is flat,
therein satisfying all the classical stability conditions as expressed by
the energy conditions. Finally, stationarity of the metric guarantees that a particle's energy is conserved.
The pathologies that plague many hypothesized metrics admitting CTCs,
e.g.\ an infinite cylinder of matter, a negative energy-distribution,
particle acceleration/blue-shifting along the CTC,
do not occur within our metric class.
%
%
\end{abstract}

\maketitle

\section{Introduction}
\label{sec:introduction}

It is well known that closed timelike curves (CTCs) are allowed solutions of general
relativity, and so time travel is theoretically possible. For years, many proposals for CTCs have been discussed
in the literature. These include van Stockum's rotating cylinder~\cite{vanStockum} (extended later by Tipler~\cite{Tipler}),
G\"{o}del's rotating universe \cite{Godel},
Wheeler's spacetime foam \cite{Wheeler},
Kerr and Kerr-Newman's region between the two horizons of the rotating black hole~\cite{Kerr},
Morris, Thorne and Yurtsever's traversable wormholes \cite{MTY},
Gott's pair of spinning cosmic strings~\cite{Gott},
Alcubierre's warp drive \cite{warp}, Ori's vacuum torus~\cite{Ori}, and some other
new proposals \cite{Gron}. All of these proposals are constructed in our observable
4D~universe. And these candidate proposals for CTCs generally suffer from at least one of the following pathologies~\cite{Visser}:
\,(1) requiring an unphysical matter distribution of infinite extent;
\,(2) violating one or more of the null, weak, strong and dominant energy conditions
(e.g., with a distribution of negative-energy),
which suggests instability of the local spacetime;
and \, (3) increasing the particle energy (``blue-shifting'') as it traverses the CTC,
thereby dissipating the source of the CTC at earlier times, leading to its non-existence.

The success of large \cite{ADD1,ADD2,ADD3} and warped \cite{RS1,RS2} extra dimensions has led many people to think of
gravitons or gauge-singlet particles taking ``shortcuts" through the extra
dimensions~\cite{shortcut1,shortcut2,shortcut3,shortcut4,shortcut5}.
For instance, a signal
such as a graviton may take a ``shortcut" from one point on the brane through the bulk and return
to the brane at a different point,
but with a transit time shorter than that for a photon traveling along a brane geodesic
between the same two points. Although the ``shortcut" allows for superluminal communication,
the path still obeys time-ordering and so does not lead to a CTC.
A shorter time-of-flight is not the same as time traveling backwards.
Using the idea of asymmetrically-warped extra-dimensions~\cite{csaki},
it has been shown that paths can be glued together to form CTCs \cite{Tom}.
But these paths are not solutions of geodesic equations,
and so would not be traversed by particles.
Also, the CTCs in~\cite{Tom} require negative-energy (``tachyonic'') matter in the bulk, violating the weak energy condition.

The purpose of this article is to propose a new framework for CTCs in one or more extra-dimensions,
which does not suffer from any of the pathologies common to previous proposals. For definiteness, we will invoke 5D~spacetime,
but our results can be readily extended to higher dimensions.

\section{The 5D Metric}

Inspired by the idea of large extra dimensions~\cite{ADD1,ADD2,ADD3} and guided by
analogy with
G\"{o}del's rotating universe~\cite{Godel} and the CTCs therein,
we are led to consider a metric off-diagonal in extra dimension
($u$, with size $L$) and time $t$.

We assume that the extra dimension is compactified.
For simplicity, we take its topology to be that of a circle (technically, a 1-sphere $S^1$).
The periodic boundary condition then requires the point $u+L$ to be identified with $u$.
With further simplicity in mind, we consider the following time-independent (``stationary'') metric:
\bea
\label{metric}
d\tau^2= \eta_{ij}dx^{i}dx^{j}+dt^2+ 2 \,g(u)\,dt\,du -h(u)\,du^2\,,
\eea
where $i,\,j=1,\,2,\,3$, and $ \eta_{ij}$ is the spatial part of the Minkowski metric.
The 4D metric induced from this 5D metric is completely Minkowskian.

This off-diagonal metric with the term $g(u)$ is reminiscent of the metric constructions
of G\"odel and van Stockum-Tipler (GvST) which admitted CTCs.
GvST imagined a physical rotating cylinder in 4D;
here the ``rotation'' is in $u$-space, with its axis of rotation parallel to the 4D~brane.
However, unlike the GvST construction with a physical 4D~cylinder of matter,
here it is the extra dimension itself which provides the cylinder.

The determinant of our metric is
$
\rm{Det}[g_{\mu\nu}] = g^2+h$.\,
The spacelike nature of the $u$ coordinate requires $\rm{Det} > 0 $ for the entire 5D metric, which, in turn,
requires that $g^2 +h >0 $ for all $u$.
It is desirable to maintain a Minkowski metric as the brane is approached.
Thus, we set $\rm{Det}(u=0) =g_0^2 +h_0=+1$, where $g_0\equiv g(0)$ and $h_0\equiv h(0)$.

The metric tensor must reflect the $S^1$ topology of the compactified extra dimension.
Thus, $g(u)$ and $h(u)$ must be periodic functions of $u$ with period $L$. We expand $g(u)$ in terms of the Fourier modes:
\bea
\label{genmetric}
g(u)&=&g_0+A  -\sum_{n=1}^\infty \left\{
   a_n\,\cos\left(\,\frac{2\pi\,n\,u}{L}\,\right)\right.\nonumber \\
   && \left. ~~~~~~~~~~~~~~~~~~~~~~~+b_n\,\sin\left(\,\frac{2\pi\,n\,u}{L}\,\right)
   \right\}\,,
\eea
where $A\equiv \sum_{n=1}^\infty a_n$ is a constant.
A similar expression can be written down for $h(u)$, but it will not be needed.
We note in passing that the value of the off-diagonal metric element $g(u)$
averaged over the path through the compact dimension is ${\bar g}=g_0 +A$;
thus, the physical meaning for $A={\bar g}-g_0$
is the deviation of mean $g$ from the brane value $g_0$.


\section{Geodesic Equations and their Solutions}
\label{sec:geodesics}

The next task is to obtain the geodesic equations of motion and solve for their solutions.
Since the metric~(\ref{metric}) is completely Minkowskian
on the brane, the geodesic equations of motion along the brane are just $\ddot{\vec r}=0$,
where the dot-derivative denotes differentiation with respect to the proper time, $\tau$.
Solutions to these geodesic equations are simply
%
\beq{dotr}
\dot{\vec r}=\dot{\vec r}_0\,,~~{\rm or}~~~{\vec r}={\vec r}_0\,\tau\,.
\eeq

The geodesic equations for $t$ and $u$ are more interesting.
Due to the time-independence (``stationarity'') of the metric,
there exists a timelike Killing vector;
the corresponding conserved quantity is
\be
\label{timeconst}
\dot{t}+g(u)\,\dot{u}=\gamma_0+g_0\,\dot{u}_0\,,
\ee
where we have evaluated the right-handed side at its initial ($\tau=0$) value.
Given this conserved quantity, it is almost evident that time will run backwards ($\dot{t}<0$), provided that
the condition $g(u)\,\dot{u}>\gamma_0+\dot{u}_0\,g_0$ is consistent with the geodesic equation for $u$.

The geodesic equation for $u$ is
\be
\label{2nd_geodesic}
2\,(g\,\ddot{t} - h\,\ddot{u}) - h'\,\dot{u}^2=0\,,
\ee
where we use the superscript ``prime" to denote differentiation with respect to $u$.
We can eliminate $\ddot{t}$ and $\ddot{u}$ from Eq.~(\ref{2nd_geodesic}).
First, we take the dot-derivative of Eq.~(\ref{timeconst}).
Then we rewrite Eqs.~(\ref{timeconst}) and (\ref{2nd_geodesic}) as
\bea
\label{geodesic_eqns}
\ddot{t}(\tau) &=&  \frac{1}{2} \,\frac{-2g' h+gh'}{g^2+h}\,\dot{u}^2 \,,\\
\label{geodesic_u}
\ddot{u}(\tau) &=& -\frac{1}{2} \,\frac{2gg'+ h'}{g^2+h} \,\dot{u}^2=-\frac{1}{2} \,\ln' (g^2+h)\,\dot{u}^2 \,.
\eea
Inspection of these two geodesic equations suggests
that we fix the determinant to be unity not just on the brane, but everywhere.
For simplicity, we do so:
\be
\label{parameterize}
\Det(u)=g^2(u)+h(u) = 1 \,,\quad\ \forall\ u\,.
\ee
Thus, $h(u)=1-g^2(u)$ everywhere.
Once the metric function $g(u)$ is given by the Fourier series of Eq.~\eqref{genmetric},
then the second metric function $h(u)$ is automatically determined.
Substituting Eq.~\eqref{parameterize} into Eq.~\eqref{geodesic_u} immediately leads to
\bea
\label{proper_velocity}
\dot{u}(\tau) &=& \dot{u}_0\,,\\
\label{u}
u(\tau)&=& \dot{u}_0\, \tau\,,~~~~~~ (\,\textrm{mod}~ L\,)\,.
\eea
The solution for $t$ can easily be obtained from Eq.~\rf{timeconst} as\\
$t(\tau)= (\gamma_0+g_0\,\dot{u}_0)\,\tau -\int^{u(\tau)}\, du\;g(u)$.
Using Eq.~\eqref{u}, we rewrite this expression in a form that
is more useful for later discussions:
\beq{intermed}
t(u)=\left( g_0 +\frac{1}{\beta_0}\right)\,u - \int_0^u du\,g(u)\,.
\eeq
Here we have introduced the symbol
$\beta_0=\frac{{\dot u}_0}{\gamma_0}=\left(\frac{du}{dt}\right)_0$\,
for the initial velocity of the particle along $u$-direction,
as measured by a stationary omniscient observer on the brane.
Analogous to those historical CTCs arising from metrics describing rotation, we will
say that a particle with $\beta_0>0$ is ``co-rotating'', while a particle with $\beta_0<0$ is ``counter-rotating''.

\section{Closed Timelike Curves}
\label{sec:CTCpossibility}

Closed timelike curves, by definition, are geodesics that return a particle to the same space coordinates from which it left,
but with a negative time so that its arrival equates to or precedes its departure.
Due to the periodic boundary condition from the $S^1$ topology of
the compactified extra dimension, a particle created on the brane but propagating into the extra dimension
will necessarily come back to the brane.
So the ``closed'' condition for a CTC is satisfied automatically by a compactified metric.
We note that when the trivial motion along the brane ${\dot{\vec r}}=$~constant
is added to the geodesic solution for $u(\tau)$, there results a helical particle motion
which periodically intersects the brane.

The ``timelike'' condition for a CTC requires that when the particle returns to the initial space coordinates,
the time elapsed as viewed by a stationary observer is zero or negative.
To ascertain whether the travel time can be negative, we must solve the geodesic equation for time,
Eq.~\eqref{intermed}.
With the general $g(u)$ given by Eq.~\eqref{genmetric},
we can perform the integration in Eq.~\eqref{intermed} to obtain
\bea
\label{tu}
t(u)&=&\left(\frac{1}{\beta_0}-A\right)\:u
   + \left(\frac{L}{2\pi}\right)\sum_{n=1}^\infty \left(\frac{1}{n}\right)
   \left\{
     a_n\,\sin\left(\frac{2\pi\,n\,u}{L}\right)\right. \nonumber \\
     && \left. ~~~~~~~~~~~~~ + b_n\,\left[1-\cos\left(\frac{2\pi\,n\,u}{L}\right)\right]
   \right\}\,.
\eea
Due to the periodic boundary condition from
the compactified extra dimension, the particle returns to the brane at $u=\pm \,N\,L,\;N=1,2,\dots$,
after traversing $N$ times around the extra dimension.
Here the $\pm$ signs hold for co-rotating and counter-rotating particles, respectively.
At the $N^{th}$ return, the time measured by a stationary clock on the brane,
as given by Eq.~\rf{tu}, will be
\beq{tNL}
t_N\equiv t(u=\pm \,N\,L)=\pm \,\left( \,\frac{1}{\beta_0}-A\,\right)\,N\,L\,.
\eeq
Interestingly, $t_N$ depends on the Fourier modes only through $A=\sum_{n=1} a_n$,
and is completely independent of the $b_n$.
Thus, the potential for a CTC arise only from the cosine modes in $g(u)$, and not the sine modes. 
In fact, we can show that a single mode from the set $\{a_n\}$ is sufficient to admit a CTC.
(This implies a necessary but not sufficient condition on $g(u)$ for the possible
existence of a CTC: for the ${\rm Det}=1$ metric, $g(\frac{L}{2})$ must differ from $g_0$.)

To have a viable CTC, we require \,$t_N <0$, or equivalently,\, $\pm\, N\,(\,\frac{1}{\beta_0}-A\,)<0$.
For a co-rotating particle, $\beta_0$ is positive, and $t_N <0$ is satisfied only if
\beq{Acondition}
A > \frac{1}{\beta_0}\,.
\eeq
For a counter-rotating particle, $\beta_0$ is negative, and $t_N <0$ is satisfied only if
\beq{Acondition2}
A,\,\beta_0 < 0 \quad{\rm and}\quad  |A|\,>\,\left|\frac{1}{\beta_0}\right|\,.
\eeq
Thus, a viable CTC requires $\textrm{sign}(A)$ to be the same as $\textrm{sign}(\beta_0)$
in either case of co-rotating or counter-rotating particles.
Once Nature chooses the constant $A$ with a definite sign,
these CTC conditions for co-rotating and counter-rotating particles are not compatible.
For definiteness in what follows, we will assume that $A>\frac{1}{\beta_0}$ is satisfied for some $\beta_0$.
Then only the co-rotating particles with sufficiently large initial bulk-velocity
can traverse the CTC to go backward in time;
the counter-rotating particles go forward in time.

Remarkably, the conditions \eqref{Acondition} and \eqref{Acondition2} can be satisfied even if $|\beta_0| < 1$.
This means that Nature does not need superluminal speeds to realize CTCs.
We exhibit the possibilities implied by our metric in Figs.~\rf{fig:regions} and~\rf{fig:worldlines}.
The geodesic may describe subluminal, superluminal, or CTC travel, depending on the
value of the positive parameter product $\beta_0\,A$.

\begin{figure*}[ht]
\includegraphics[width=9.5cm]{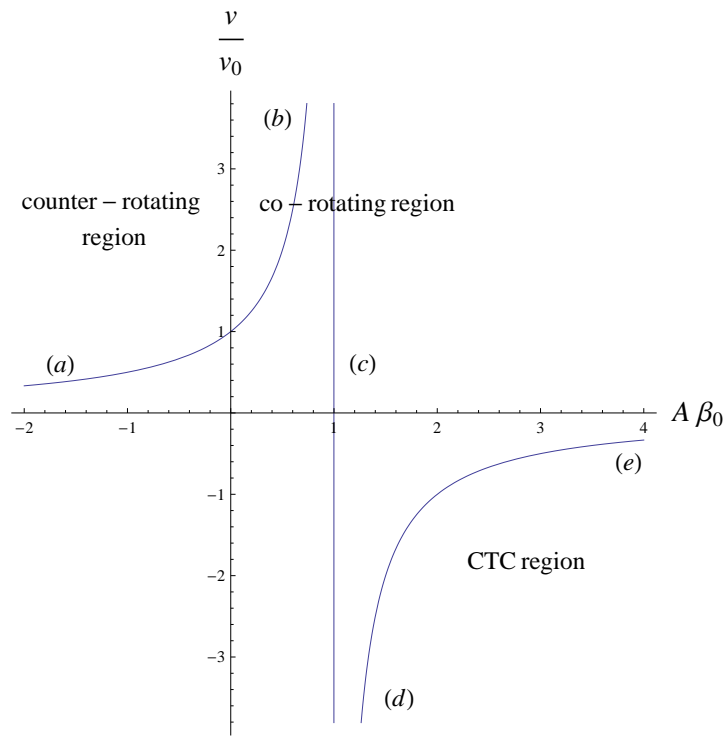}
\caption{
The ratio of apparent brane velocity $v$ to initial brane velocity $v_0$ is plotted against
$\beta_0\:A$. The co-rotating particle can move superluminally in either time direction, while
the counter-rotating particle always moves subluminally forward in time.
Note that when the lightcone crosses the horizontal axis of the spacetime diagram at $\beta_0\:A = 1$,
the brane velocities are divergent.
For $\beta_0 A > 1$, the co-rotating geodesic corresponds to a CTC.
The regions denoted by (a), (b), (c), (d) and (e) correspond to the worldlines with the same labels
in Fig.~\ref{fig:worldlines}.
\label{fig:regions}}
\end{figure*}
\begin{figure*}[ht]
\includegraphics[width=8cm]{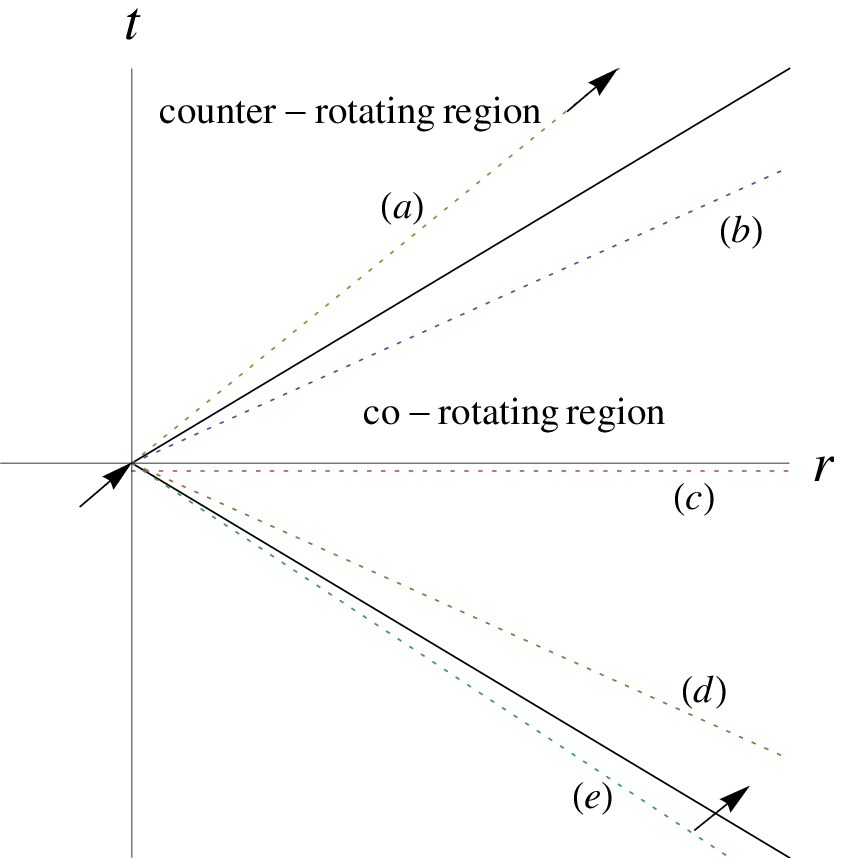}
\caption{The worldlines denoted by (a), (b), (c), (d) and (e) correspond to the
regions with the same labels in Fig.~\ref{fig:regions}. For instance, the worldline (a) corresponds to a counter-rotating particle traveling
(forward in brane/coordinate time $t$) within the forward-light-cone. The worldlines
(c), (d) and (e) correspond to a co-rotating particle traveling beyond the brane's forward-light-cone.
For the wordline (b), the particle travels superluminally but moves forward in brane time.
The worldline (c) is horizontal and so the particle is ``moving" instantaneously in brane time.
The worldlines (d) and (e) represent particles traveling with superluminal and subluminal velocities respectively, and in both cases,
the particles travel backwards in brane time (signifying a CTC).  For all worldlines, the dashed nature of the line represents the
stroboscopic piercings of the brane by the particle circumscribing the compactified extra-dimension.
\label{fig:worldlines}}
\end{figure*}

\section{World-line Analysis}
\label{sec:light-cone}

In this section, we generalize the analysis of massless particle light-cones
in Ref.~\cite{Tom} to massive particles, and discuss the result.

The realization of CTCs requires that the particle world-line tips so that
the motion in the $u$-direction occurs backwards in time $t$ as measured from the brane.
In terms of the spacetime slope $s\equiv dt/du$ for the world-line along the $\pm u$-directions,
the line element in Eq.~\rf{metric} is simply
\beq{slopes1}
\frac{d\tau^2}{du^2}= \frac{1}{{\dot u}^2(u)}=s^2 + 2s\,g(u) -h(u)\,,
\eeq
where we have ignored a possible nonzero velocity $d{\vec r}/dt$ along the brane
which would not affect the following discussions.
The solutions to the above quadratic equation are
\be
\label{slopes2}
s_\pm (u) = -g(u)\pm\sqrt{g^2(u)+h(u)+\frac{1}{\dot{u}^2(u)}}\,,
\ee
where $s_+$ and $s_-$ represent the slopes for the co-rotating and counter-rotating world-lines respectively.
The consistency of this assignment can be checked by noting that
$s(u=0)$ is just~$\gamma_0/{\dot u}_0$, which implies that $\textrm{sign}(s(0))=\textrm{sign}({\dot u}_0)$.

It is more illuminating to put Eq.~\rf{slopes2} into the form
\bea
\label{sumNproduct}
s_-(u) \,+ \,s_+(u) &=& -2\,g(u)\,, \\
s_-(u)\;s_+ (u) &=& -\left(\,h(u)+\frac{1}{\dot{u}^2(u)}\,\right)\,.
\eea
In conventional 4D Minkowski space, $g_0$ would be zero and $h_0$ would equal unity.
For our CTC in 5D, it is crucial that $g_0\ne 0$,
but we assume a semblance of the Minkowski space limit by taking $h_0\ge 0$.
The world-line of a CTC must tip into the negative $t$ region, and so its slope must pass through zero.
This means that $s_- (u)\; s_+ (u)$, or equivalently, $h(u)+\dot{u}^{-2}$, must pass through zero.
Suppose that this happens at $u=\ustar$; then $h(\ustar)=-{\dot\ustar}^{-2}$.
With our previous assumption that $g^2+h=1$ everywhere, we have
$g^2(\ustar)=1-h(\ustar)=1+\dot{u}^{-2}$.  But by Eq.~\rf{proper_velocity},
we may write this result as $|g(\ustar)|=\sqrt{1+\dot{u}_0^{-2}}$.
On the other hand, Eq.~\rf{proper_velocity} and the constraint $h_0\ge 0$ imply that
$|g_0|<1$.
%
%
We learn that time will become negative if $|g(u)|$ rises from its value
$|g_0|\le 1$ on the brane to above $\sqrt{1+\dot{u}_0^{-2}}$ in the bulk.
For a general metric function of Eq.~(\ref{genmetric}), this requirement on $g(u)$ is easily accommodated.

We reinforce two earlier lessons from this world-line analysis.
The first is that for a given sign of $A$, only one of the co-rotating and counter-rotating particles can experience the CTCs.
The reason here is that only {\sl one} edge of the world-line can tip below the horizontal axis into the negative-time half-plane.
The second lesson is that since time begins to flow backwards only after $|g(u)|$ rises to $\sqrt{1+\dot{u}_0^{-2}}$,
there may exist a critical $({\dot u}_0)_{\rm min}$ (equivalent to a minimum $\beta_0$)
below which CTCs are not accessible.

\section{Resemblance with 4D Spinning Strings}
\label{sec:compared2spinning}

Our class of 5D metrics admitting CTCs resembles in some ways the well-studied metric
for a 4D spinning cosmic string \cite{DJtH-Annals,DJ-Feinberg}:
\beq{spinstr}
d\tau^2_{\rm\stackrel{spinning}{string}}= (dt+ 4\,G\,J \,d\theta)^2-dr^2-(1-4\,G\,m)^2\,r^2\,d\theta^2-dz^2\,,
\eeq
where $G$ is Newton's constant, $J$ is the angular momentum,
and $m$ is the mass per unit length of the cosmic string.

In three spacetime dimensions, the Weyl tensor vanishes, and so any region without a gravitational source must be flat.
Consequently, in the region outside the spinning string, the local Minkowski coordinates may be extended to
cover the entire region. In particular, one could change the coordinates in Eq.~\rf{spinstr}
to $\tilde{t} = t+ 4\,G\,J \,\theta $ and\, $\varphi =(1-4\,G\,m)\,\theta $ such that
the metric becomes Minkowskian, with the conformal factor being unity.
Similar to $\theta$, \,$\varphi$ is periodic and is subject to the identification
$\varphi \sim \varphi + 2\pi- 8\pi\,G\,m$.
It is well-known that the wedge $\Delta\varphi=8\pi \,G\,m$ should be
removed from the plane, leaving behind a cone.
While these coordinate transformations apparently lead to simplicity, in fact
$\tilde{t}$ is a pathological coordinate.
It is a linear combination of a non-compact variable $t$ and a compact variable $\theta$.
For a fixed $\theta$ (or $\varphi$), \,$\tilde{t}$ is a smooth and continuous variable.
But for a fixed $t$, one needs the identification $\tilde{t} \sim \tilde{t}+ 8\pi\,G\,J$
to avoid a ``jump'' in the new variable.
A a result, the singularity at $g_{\theta\theta}=0$, which occurs at $r=4GJ/(1-4Gm)$,
is in effect encoded in the pathological coordinate $\tilde{t}$~\cite{DJtH-Annals}.

In the $(t,\ u)$-plane, our metric has the form
\beq{ut_form}
d\tau^2 = (\,dt+g(u)\,du\,)^2 - du^2\,,
\eeq
where we have used the simplifying condition in Eq. \eqref{parameterize}. This appears similar to the 4D spinning-string metric.
Analogously, we can define a new exact differential $d\bar{t}\equiv dt+g(u)\,du$ to put our metric into the diagonal ``Minkowskian'' form:
\bea
\label{metric2}
d\tau^2 = \eta_{ij}\,dx^{i}dx^{j}+ d\bar{t}^2 - du^2\,.
\eea
This nontrivial coordinate transformation defines a new time variable $\bar{t}= t+ \int_0^{u(t)}\,du\;g(u)$\,
which is measured in the frame that ``co-rotates" with the circle $S^1$.
Since the equivalent metric is locally Minkowskian everywhere, the entire 5D spacetime is flat.
This is consistent with the theorem which states that
any two-dimensional (pseudo) Riemannian metric, whether in a source-free region or not,
is conformal to a Minkowski metric.
(Here, our $(t,\,u)$ submanifold is not only conformally flat, but has a conformal factor of unity.)
However, similar to the case of the spinning string, the topology of our 5D spacetime is non-trivial.
The new time variable $\bar{t}$ is an ill-defined variable,
a pathological combination of a non-compact $t$ coordinate and a compact $u$ coordinate.
On the brane, this time is $\bar{t}=t \pm NL\bar{g}$, where $\bar{g}\equiv \frac{1}{L}\,\int_0^L g(u)du$,
and $N$ is the number winding number of the co- or counter-rotating off-brane particle
(with the plus sign for co- and minus sign for counter-rotating).
We see that the compact nature of the $u$-dimension reveals itself in this new time coordinate in two ways:
in a ``memory'' of the winding number $N$ of the off-brane particle,
and in a knowledge of the metric element $g(u)$ averaged over the compact dimension.
(Both non-local features are necessary if the $(\bar{t},u)$-framework is to
reproduce the brane-piercing intervals given in Eq.~\rf{tNL} for the $(t,u)$-framework,
and displayed in Fig.~\rf{fig:worldlines}.)     .

We remark that the time measured by an observer (or experiment) on our brane should just be
given by $t$. The reason is that the constraint equation that reduces the 5D metric to the induced 4D metric
is simply $u(x^\mu)=0$, and taking the differential gives $du=0$.
When the latter result is substituted into the 5D metric in Eq.~\rf{metric},
the standard 4D Minkowski metric with time $t$ is induced.

\section{Pathologies of Moving 4D Cosmic Strings}
\label{sec:4Dpathology}

It has been shown by Deser, Jackiw, and 't~Hooft that the metric for the 4D spinning string
leads to CTCs~\cite{DJtH-Annals}.
However, this metric has also been criticized by themselves and others.
Their criticism is that the definition of spin becomes singular
as one approaches the string's center at $r=0$.
In contrast, this problem is absent in our compactified 5D metric of Eq.~\eqref{metric}
because there is simply no ``$r=0$'' in the $u$-space.
The ``center'' of the periodic $u$-space is not part of the spacetime.

Gott proposed an improved stringy CTC by making use of a pair of infinitely-long cosmic strings
with a relative velocity~\cite{Gott}.
In his improved scheme, the singular spin angular-momentum of a single spinning string is replaced by
the non-singular orbital angular-momentum of a two-string system.
Since each of the cosmic strings is infinitely long,
the configuration is translationally invariant along the $z$ direction,
and one can freeze the $z$ coordinate and reduce the problem to
an effective (2+1)~dimensional spacetime.
In this (2+1)D spacetime, the piercings of the two strings appear as moving dots.
Gott showed that there exists a `figure-eight'' CTC geodesic
encircling the dots and crossing between them.

However, the non-trivial topology in Gott's spacetime results in non-linear energy-momentum addition rules.
While each of the spinning cosmic strings carries a timelike energy-momentum vector,
the two-string center-of-mass energy-momentum vector
turns out to be spacelike or tachyonic~\cite{Hooft,Shore}.
Even though Gott's CTC does not violate the weak energy condition,
the tachyonic total energy-momentum vector leads to
violations of the null, strong and dominant energy conditions~\cite{Shore,Carroll,Tye}.
(Energy conditions are briefly discussed in the next section.)

In addition, it has been proved that in an open universe,
an infinite amount of energy is required to form Gott's CTC~\cite{Carroll}.
A related argument against the stability of Gott's CTC is the blue-shifting
of the particle traversing the CTC~\cite{Tye}.
Since the particle can traverse the CTC infinitely many times, it can be infinitely blue-shifted,
while maintaining the elapsed time as negative~\cite{Hawking,Carroll}.
This implies that the total energy of the pair of cosmic strings would have been
infinitely dissipated even before the particle enters the CTC for the first time.
The simple interpretation is that the CTC simply cannot be formed in the first place.

\section{Compactified 5D CTCs without Pathologies}
\label{sec:nopathology}

In the previous section, we discussed the pathologies of a CTC realized by
a pair of moving cosmic strings.
In fact, these pathologies are common among 4D metrics admitting CTCs.
In contrast, we will show in this section that
our compactified 5D CTCs do not suffer from any of these pathologies.

Firstly, the realization of our compactified 5D CTCs with
a compactified extra dimension only requires a flat spacetime metric.
No matter distribution, negative or positive, infinite or even finite extent, is needed.
One can easily verify that all the components of the 5D curvature tensor $R_{ABCD}$
and Ricci tensor $R_{AB}$,
derived from the metric of Eq.~(\ref{metric}), are identically zero.
Thus, by the Einstein field equation, the energy-momentum tensor $T_{AB}$ is also vanishing.
This implies that our 5D spacetime, which admits CTCs, automatically satisfies all of the
standard null, weak, strong and dominant energy conditions:
\bea
\textrm{NEC:} && T_{AB}\, l^{A}\,l^{B} \geq 0 \,, \nonumber\\
\textrm{WEC:}  && T_{AB}\, t^{A}\,t^{B} \geq 0 \,, \nonumber\\
\textrm{SEC:}  && T_{AB}\, t^{A}\,t^{B} \geq \frac{1}{2}\,
     T^{A}_{A}\, t^{B}\,t_{B} \,,   \nonumber\\
\textrm{DEC:}  && T_{AB}\, t^{A}\,t^{B} \geq 0~~
     \textrm{and} ~~T_{AB}\, T^{B}_{C} \,t^{A}\,t^{C} \leq 0\,, \nonumber
\eea
where $l^{A}$ and ${t^{A}}$ are any null vectors and timelike vectors respectively.

Secondly, particles traversing the compactified 5D CTCs are not blue-shifted. This can be understood
as follows. The contravariant momentum is defined as
$p^A \equiv m\,(\dot{t},\,\dot{\vec{r}},\,\dot{u})$,
with $m$ being the mass of the particle. Correspondingly, the covariant five-momentum is given by
\beq{PdnA}
p_A = G_{AB}\,p^B = m\,\left(\,\dot{t}+ g \,\dot{u},\;-\dot{\vec{r}},\;g\,\dot{t}-h\,\dot{u}\,\right)\,.
\eeq
From Eq.~\rf{timeconst}, it is clear that the quantity $p_0= m \,(\dot{t}+ g \,\dot{u})$ is covariantly conserved
along the geodesic on and off the brane, a result of the time-independence of the metric $G_{AB}$.
We can therefore identify this conserved quantity as the energy $E$ of the time-traveling particle.
With the energy $E$ covariantly conserved, we conclude that
the particle is not blue-shifted, and there is no dissipation of the CTC.

\section{Conclusions}
\label{sec:conclusion}

We have constructed a class of CTCs that are physical and classically stable.
Since it is the compactified extra dimension that enables the CTCs,
only the Kaluza-Klein (KK) modes of quanta can traverse through these CTCs and go backwards in time.
Wherever there is spacetime, there will be gravitons from the quantization of the metric fluctuation.
Therefore, KK gravitons will certainly be time-traveling particles, provided that our specific metric
in Eq.~\eqref{metric} is realized by Nature.
If Standard Model particles are confined to
our familiar 4D brane as in the framework of large extra dimensions~\cite{ADD1,ADD2,ADD3},
we may anticipate that the KK modes of gauge-singlets (e.g.\ Higgs singlets or sterile neutrinos)
could also be CTC time-travelers.

Finally, we mention that our derivation has been purely classical.
Whether or not our results survive in a quantum mechanical picture is another story,
yet to be written..

\acknowledgments

C.M. Ho and T.J. Weiler were supported in part by the
Department of Energy grant DE-FG05-85ER40226.

{}

\end{document}